%Paper: hep-ph/9404320
%From: PARANJ@LPS.UMONTREAL.CA
%Date: Thu, 21 Apr 1994 15:25:46 -0400 (EDT)

\input phyzzx
\tolerance=5000
\sequentialequations
\overfullrule=0pt
\nopubblock
\PHYSREV
\nopagenumbers
\vsize8.5truein\voffset.25truein
\hsize6truein\hoffset.25truein

\def\entier{{\rm Z}\mskip-6mu{\rm Z}}

\def\reel{{\rm I}\mskip-3.5mu{\rm R}}
\def\haf#1{{{#1}\over 2}}

\def\sla{\raise.16ex\hbox{$/$}\kern-.57em}
\def\Dsl{\kern.2em\raise.16ex\hbox{$/$}\kern-.77em\hbox{$D$}}

\def\Asl{\kern.2em\raise.16ex\hbox{$/$}\kern-.77em\hbox{$A$}}
\def\Bsl{\kern.2em\raise.16ex\hbox{$/$}\kern-.77em\hbox{$B$}}
\def\square{\kern1pt\vbox{\hrule height 1.2pt\hbox{\vrule width 1.2pt\hskip 3pt
   \vbox{\vskip 6pt}\hskip 3pt\vrule width 0.6pt}\hrule height 0.6pt}\kern1pt}

\gdef\journal#1, #2, #3, 1#4#5#6{               % Journal reference.  Comma
%%sets
    {\sl #1~}{\bf #2}, #3 (1#4#5#6)}            % off: name, vol, page, year

\def\np{\journal Nucl. Phys., }

\def\prl{\journal Phys. Rev. Lett., }

\def\prd{\journal Phys. Rev. D, }

\def\cmp{\journal Comm. Math. Phys., }

%\line{\hfill UdeM-LPN-TH-xxx}
%
\centerline{\bf Low Energy Skyrmion-Skyrmion Scattering}
\bigskip\bigskip
\centerline{ T. Gisiger and M. B. Paranjape}
\medskip
\vskip.2cm\singlespace
\centerline{{\it Laboratoire de Physique Nucl\'eaire,
Universit\'e de Montr\'eal, C.P. centreville}}
\centerline{{\it Montr\'eal, Qu\'ebec, Canada H3C 3J7}}

\bigskip
\centerline{ABSTRACT}
\tenpoint\baselineskip12pt
\narrower{\smallskip\noindent
We study the scattering of Skyrmions at low energy and large separation using
the method proposed by Manton of truncation to a finite number of degrees
freedom. We calculate the induced metric on the manifold of the union of
gradient flow curves, which for large separation, to first non-trivial order is
parametrized by the variables of the product ansatz.
\smallskip}

\bigskip
\twelvepoint\baselineskip14pt

\REF\man{ N.S. Manton, \prl  60, 1916, 1988.}

\REF\gibman{ G.W. Gibbons, N. S. Manton, \np B274, 183, 1986.}

\REF\sam{T.M. Samols, \cmp 145, 149, 1992.}

\REF\sky{T.H.R. Skyrme, \journal Proc. Roy. Soc. Lon., 260, 127, 1961.}

\REF\wit{ E. Witten, \np B223, {422, 433}, 1983.}

\REF\kutpet{M. Kutschera, C.J. Pethick, \np A440, 670, 1985.}

\REF\islletpar{K. Isler, J. LeTourneux and M.B. Paranjape,
		\prd 43, 1366, 1991, and references herein.}

\REF\walwam{T.S. Walhout and J. Wambach, \prl 67, 314, 1991.}

\REF\sch{ B. Schroers, Durham preprint, DTP 93-29, hep-ph/9308236.}

\REF\adknapwit{G.S. Adkins, C.R. Nappi and E. Witten, \np B228, 552, 1983.}

The calculation of the scattering of baryons is an intractable problem in
Q.C.D..  Some progress can be made using the low-energy effective, field
theoretic description.  The Skyrme model[\sky] corresponds to the effective
degrees of freedom of low energy Q.C.D..  The Skyrme model is described by the
Lagrangean,
$${\cal L} =
{f_\pi^2 \over 4} tr(U^\dagger \partial_\mu U U^\dagger \partial^\mu U)
+ {1\over 32 e^2}\,tr( [U^\dagger \partial_\mu U,U^\dagger \partial_\nu U]^2)
$$
where $U(x)$ is a unitary matrix valued field. We take
$$
U(x) \in SU(2).
$$
The Skyrme Lagrangean contains the first terms of a systematic expansion in
derivatives of the effective Lagrangean describing low energy
interaction of pions. It is derivable from QCD hence $f_\pi$ and $e$ are in
principle calculable parameters. It includes spontaneous breaking of chiral
symmetry
$$
\eqalign{
&SU_L(2)\times SU_R(2)\rightarrow SU_V(2)\quad{\rm with}\cr
&U(x)=1+i\vec\pi (x)\cdot\vec\tau\cdots\quad{\rm where}\cr
&\vec\pi (x)\sim\quad{\rm pions}.}
$$
What is even more surprising is that it includes the baryons as well.  They
arise
as topologically stable, solitonic solutions of the equations of motion. The
original proposal of this by Skyrme [\sky] in the 60's was put on solid
footing by
Witten [\wit] in the 80's.  The solitons, which are called Skyrmions,
correspond to
non-trivial mappings of $\reel^3$ plus the point at infinity into $SU(2)$:
$$U(x): \reel^3 + \infty \to SU(2) = S^3.$$
But
$$\reel^3 + \infty = S^3$$
thus the homotopy classes of mappings
$$U(x): S^3 \to S^3$$
which define
$$\Pi_3(S^3) = \entier$$
characterize the space of configurations.  The topological charge of each
sector
is given by
$$
N = {1 \over 24 \pi^2} \int{d^3 \vec x\, \epsilon^{ijk}
tr(U^\dagger \partial_i U U^\dagger \partial_j U U^\dagger \partial_k U)}
$$
which is identified with the baryon number.  The Skyrmion solution is given by
the configuration
$$
U_S(\vec x)=e^{if(\left|\vec x\right|)\hat x\cdot\vec\tau}
$$
where $f(\left|\vec x\right|)$ is a decreasing function which starts at $\pi$
at the origin and achieves $0$ at $\left|\vec x\right| =\infty$,
asymptotically varying as
$\kappa\over \left|\vec x\right|^2$.  A Skyrmion at position $\vec R$ with
orientation $A\in SU(2)$ corresponds to the configuration
$$
U(A,\vec R,\vec x)=A U_S(\vec x-\vec R)A^\dagger.
$$
Quantum states of definite momentum, isospin and spin obtained by
quantizing  $\vec R$ and $A$ correspond to the baryons; the nucleons, deltas,
etc..
$$
\vec J=\vec L+\vec T
$$
is conserved, however,
$$
\left(\vec L\right)^2=\left(\vec T\right)^2.
$$
The nucleons correspond to
$$
\left|\vec L\right|=\left|\vec T\right|=\haf 1.
$$

With the input of two parameters, $f_\pi$ and $e$, everything else can be
predicted for the baryons. Agreement with experiment is within $10\%\sim 30\%$
for $M_N, M_\Delta -M_N ,<r^2>^\haf 1_{T=0,1}, \mu_p ,\mu_n ,g_A, g_{\pi
NN}\cdots $[\adknapwit ].

The sector with $B=2$, should contain the deuteron as a bound state of minimal
energy and the scattering of two nucleons.  Even the classical scattering is
too
difficult to compute.  There are an infinite number of degrees of freedom.
One has to in principle solve a non-linear partial differential equation of
motion for the time evolution.

An idea put forward by Manton[\man] was to look
for an appropriate truncation of the degrees of freedom.  He first considers
the
case of theories of the Bogomolnyi type, those theories which admit {\it
static}
soliton solutions, usually in the topological two soliton sector, which
asymptotically describe two single solitons at arbitrary positions and relative
orientations. The configuration at small separation contains, in general,
strong
deformations of the individual solitons and in fact they lose their identity.
The set of configurations have, however, the same energy since they correspond
to the continuous variation of a finite number of parameters, the modulii.
Otherwise they could not be  stationary points of the potential.
In general, for solitons corresponding to a topological quantum number, the
modulii space corresponds to the sub-manifold of minimum energy configurations
within the given topological sector. Manton suggests that the low energy
scattering of solitons, with initial configuration on this sub-manifold
corresponding to asymptotic, single solitons, with arbitrarily small initial
velocity tangent to the sub-manifold, will self-consistently be constrained
to remain on the sub-manifold.  Since the potential energy is a constant on the
sub-manifold the resulting dynamics reduces to geodesic motion on the
sub-manifold in the induced metric on the sub-manifold from the kinetic term.
It is a difficult task to prove such a truncation of degrees of freedom
in a mathematically rigorous fashion, however, it does seem intuitively
correct. The non-linearity of the theory implies the coupling of the degrees of
freedom corresponding to the sub-manifold with all other excitations through
the potential. We are assuming that these are negligible. Manton and
Gibbons [\gibman] applied this program with remarkable success to the case of
magnetic monopoles in the BPS limit and it has also been applied to vortex
scattering in a similar limit [\sam].

The generalization to the more common situation where the set of static
solutions correspond to a finite set of critical points proceeds as follows.
The critical points are typically a minimum energy configuration which is
essentially a bound state of two solitons, an asymptotic critical point which
corresponds to two infinitely separated solitons and possibly a number of
unstable non-minimal critical points of varying energies of the same order.
These critical points are degenerate with a finite number of degrees of
freedom. They are connected by special paths, the paths of steepest
descent or equivalently the
gradient flow curves. In this case Manton proposes that the dynamics will be
constrained to lie on the sub-manifold comprising of the union of all these
curves. This again is intuitively reasonable. If we think of the space of all
configurations as a large bag, the bottom surface of the bag will correspond to
this sub-manifold, and a slow moving marble rolling on the bottom will tend to
stay there.

The Skyrme model falls into the second case.  We identify the corresponding
sub-manifold for well-separated Skyrmions and we calculate the induced metric
to lowest non-trivial inverse order in the separation from the kinetic term.
This is the first step towards calculating the scattering of Skyrmions
in this formalism.

Thus for the scattering of two
Skyrmions, we are looking at the sector of baryon number equal to 2.
In this sector the minimum energy configuration should correspond to the bound
state of two Skyrmions, which must represent the deuteron. The asymptotic
critical point corresponds to two infinitely separated Skyrmions.
There exist, known, non-minimal critical points, corresponding to a spherically
symmetric configuration, the di-baryon solution [\kutpet]. The energy of this
configuration is about three times the energy of a single Skyrmion. There
are also, possibly, other non-minimal critical points with energy less than two
infinitely separated Skyrmions [\islletpar]. The scattering of two
Skyrmions will take place
on the union of the paths of steepest descent which connect the various
critical points.

We consider the scattering only for large separation. In this way we do not
have to know the structure of this manifold in the complicated
region where the two Skyrmions interact strongly and consequently are much
deformed. In the region of large separation the product ansatz corresponds to
$$
\eqalign{
U(\vec x) &= U_1(\vec x - \vec R_1) U_2(\vec x - \vec R_2)\cr
          &= AU(\vec x - \vec R_1)A^\dagger B
 U(\vec x - \vec R_2)B^\dagger}
$$
where $U(\vec x - \vec R_1)$ and $U(\vec x - \vec R_2)$ correspond to the field
of a single Skyrmion solution centered at
$R_1$ and $R_2$ respectively. The full Skyrme model dynamics implies a
deformation of each Skyrmion. We will neglect this deformation.

It remains to calculate the metric on the sub-manifold parametrized by the
product ansatz. We replace $\vec R_1-\vec R_2$ by $\vec d$ placing us in the
center of mass reference frame and reducing the number of degrees of freedom of
the system to nine.  We find, along with Schroers[\sch] the interesting
result that the metric behaves like $1/d$ where $d=|\vec d|$ is the
separation. We find the kinetic energy:
$$
\eqalign{
T =& -2 M + {1\over 4} M \dot{\vec d^{\;2}} + 2 \Lambda \bigl({\cal L}^a(A)\,
{\cal L}^a(A) + {\cal L}^a(B)\,{\cal L}^a(B)\bigr)\cr
& + {\Delta\over d} \epsilon^{iac}\epsilon^{jbd}\;{\cal R}^c(A)\,{\cal
R}^d(B)\;
\bigl(\delta^{ij}-\hat{d}^i \hat{d}^j\bigr)\, D_{ab}(A^\dagger B) +O(1/d^2).}
$$
where
$$
\eqalign{M =\;& 4\pi \int_0^\infty r^2dr\cr
\times &{ \biggl\{ {1\over 8} f_\pi^2 \biggl[\biggl({\partial F\over \partial
r}\biggr)^2\!\!+ 2\,{\sin^2 F\over r^2}\biggr]+{1\over 2 e^2}{\sin^2 F\over
r^2}
\biggl [{\sin^2 F\over r^2} + 2\biggl({\partial F\over \partial
r}\biggr)^2\biggr] \biggr\} }\cr}
$$
is the mass of a Skyrmion and
$$\Lambda = (ef_\pi )^3\int{r^2 dr \sin^2 F\biggl[ 1+ {4\over (ef_\pi)^2 }
\biggl(F'^2+ {\sin^2 F\over r^2}\biggr)\biggr] }
$$
is its inertia momentum and where $\Delta=2 \pi \kappa^2 f_\pi^2$, $F(r)\sim
\kappa/r^2$ at large $r$, ${\cal R}^a (A)\equiv {\cal R}^a_0(A)$ and $\hat d =
\vec d/d$. The metric can be easily obtained from this expression by
choosing local coordinates on the product ansatz manifold and extracting the
quadratic form relating their time derivatives.

The potential [\islletpar] between two Skyrmions can be calculated to give
$$
V= 2\Delta {(1-\cos\theta)(3 \,(\hat n\cdot\hat d)^2 -1)
\over d^3}
$$
where $\theta$, $\hat n$ pick out the element of $SU(2)$ given by $A^\dagger
B$.
The potential is clearly of higher order than the metric, hence the
dominant contribution to the scattering at large separation comes only from the
metric. Thus to leading order we may even neglect the potential and then the
problem reduces to calculating the geodesics on the product ansatz manifold.

In summary, we underline the salient points of our treatment.  In principle, we
begin with QCD, which implies the Skyrme model as a low energy effective field
theory.  The classical soliton dynamics of the Skyrme model, can be harnessed
using the truncation to a finite number of relevant degrees of freedom.
The ensuing dynamics for the collective coordinates reduces to geodesic
motion on the manifold parametrized by the product ansatz.  Our treatment
requires no ad hoc parameters and each approximation has a well defined
domain of validity.  We are presently working out the details of projecting
onto semi-classically quantized nucleonic states and the implications for
nucleon-nucleon scattering.
\vfill
\break
\vskip 2.0cm
\centerline{\bf Acknowlegements}
\vskip 1.0cm
We thank M. Temple-Raston for useful discussions. This work supported in part
by NSERC of Canada and FCAR of Qu\'ebec.

\refout

\bye